# Structural transition and uranium valence change in $UTe_2$ at high pressure revealed by x-ray diffraction and spectroscopy

Yuhang Deng[1†], Eric Lee-Wong[1†], Camilla M. Moir[1], Ravhi S. Kumar[2], Nathan Swedan[1], Changyong Park[3], Dmitry Yu Popov[3], Yuming Xiao[3], Paul Chow[3], Ryan E. Baumbach[4,5], Russell J. Hemley[6], M. Brian Maple[1*]

[1]Department of Physics, University of California, San Diego, CA 92093, USA

[2]Department of Physics, University of Illinois Chicago, Chicago, IL 60607, USA

[3]High Pressure Collaborative Access Team, X-ray Science Division, Argonne National Laboratory, Argonne, IL 60439, USA

[4]National High Magnetic Field Laboratory, Florida State University, Tallahassee, FL 32306, USA

[5]Department of Physics, Florida State University, Tallahassee, FL 32306, USA

[6]Departments of Physics, Chemistry, and Earth and Environmental Sciences, University of Illinois Chicago, Chicago, IL 60607, USA

†These authors contributed equally to the work.

*Corresponding author: M. Brian Maple. Email: mbmaple@ucsd.edu

## Abstract

High pressure x-ray diffraction up to 30 GPa in conjunction with resonant emission x-ray spectroscopy and partial fluorescence yield x-ray absorption spectroscopy up to 52 GPa were used to study how the structural and electronic properties of $UTe_2$ evolve with pressure at room temperature. An orthorhombic to tetragonal phase transition was observed to occur between 5 and 7 GPa, with a large volume collapse of nearly 11% and a nearest U-U distance increase by about 4%. This lower to higher symmetry transition suggests less 5*f* electron participation in bonding when the weakly correlated superconducting phase in the tetragonal structure of $UTe_2$ appears. Beyond 7 GPa, no new structural transitions were found up to 30 GPa. The resonant x-ray emission spectra clearly demonstrate an intermediate valence of U, nearly +3.74 at 1.8 GPa and room temperature, and reveal that the U valence shifts towards 4+, passes through a peak at 2.8 GPa, and then decreases towards 3+ and settles to a nearly constant value above 15 GPa. These experiments reveal that some fundamental structural and valence changes occur in $UTe_2$ at relatively low pressures, which could be responsible for the interplay between unconventional superconductivity, magnetic ordering, and weakly correlated superconductivity that is manifested in the temperature-pressure phase diagram of $UTe_2$.

## Introduction

The heavy fermion *f*-electron superconductor $UTe_2$ has attracted a great deal of attention, driven by an interest in developing a fundamental understanding of its extraordinary unconventional superconducting properties and the possibility that it exhibits spin-triplet superconductivity with potential applications in quantum computation [1, 2]. The compound $UTe_2$ has an enormous re-entrant upper critical field $H_{c2}(T)$ of the order of 40 T, considering its low superconducting critical temperature $T_c$ of only 2 K. In addition, there is a pocket of high magnetic field superconductivity (so-called Lazarus phase) which occurs at magnetic fields $B$ between 40 T and 60 T and at angles $\theta$ between 23º and 45º, where $\theta$ is measured with respect to the $b$ axis in the $b$-$c$ plane of the $UTe_2$ body-centered orthorhombic unit cell [3, 4].

Our approach to study the electronic properties of $UTe_2$ uses pressure (*P*) to tune electronic interactions in $UTe_2$. Pressure (*P*) is a clean and powerful parameter for tuning electronic interactions in quantum materials, often resulting in dramatic changes in physical properties in the vicinity of a critical pressure $P_c$ at which a second order phase transition has been suppressed to 0 K, referred to as a quantum critical point (QCP). Remarkably, unconventional forms of superconductivity and exotic magnetic phases are often found in the vicinity of the QCP. We applied two different spectroscopic techniques, U $L_3$ edge Partial Fluorescence Yield X-ray Absorption Spectroscopy (PFY-XAS) and Resonant X-ray Emission Spectroscopy (RXES), to explore the U valence change and the degree of 5*f* electrons delocalization in $UTe_2$ under high pressure. The U $L_3$ edge was chosen because of the better energy resolution from the longer 3*d* core-hole lifetime (~4 eV) compared to the 2*p* core-hole lifetime (~8 eV) [5]. The $L_{\alpha 1}$ emission from the 3*d* to 2*p* transition was found to split due to different screening effects on the 2*p* core hole by the 5*f*-electron in different configurations [6, 7], pointing unambiguously to the multiconfigurational features of U in $UTe_2$. X-ray Diffraction (XRD) measurements under high *P* were also used to determine the *P*-dependence of the lattice parameters *a*, *b*, and *c*, and volume of the body-centered orthorhombic unit cell of $UTe_2$, and search for possible crystallographic phase transitions. Measurements were taken at room temperature (RT) up to 30 GPa for XRD and up to 52 GPa for PFY-XAS and RXES.

From XRD data, we observed a structural phase transition from the body-centered orthorhombic structure (space group *Immm*) to the body-centered tetragonal structure (space group *I4/mmm*) in the range of 5 – 7 GPa, confirming the results in Refs. [8, 9] at ~5 GPa and ~4 GPa respectively. A nonmonotonic change in the U $L_3$ white-line position was seen in the PFY-XAS spectra, which indicates a pressure-induced change in the U valence or a change in the degree of localization of the 5*f* electrons. From RXES measurements, we found evidence for an initial increase in the U valence toward 4+ up to 2.8 GPa, in partial agreement with the XANES measurements in Ref. [10] and the XANES and X-ray magnetic circular dichroism (XMCD) measurements in Ref. [11]. Strikingly, with increasing pressure beyond 2.8 GPa, the U valence drops toward 3+ until ~15 GPa where it remains stable up to 52 GPa. The unusual U valence change and the appearance of the high-pressure tetragonal phase harboring a new superconducting state [9] may be related and warrants further investigation.

## Experimental Methods

$UTe_2$ single crystals were grown by chemical vapor transport and then powdered for measurements. Two batches of samples (S1 and S2) were measured in this work. Batch S1 was synthesized using the following procedure: uranium and tellurium in a 2:3 atomic ratio were sealed in a quartz tube with 3 $mg/cm^3$ of iodine and kept at a temperature gradient of 1060 $^{\circ}C$ at the hot end and 1000 $^{\circ}C$ at the cold end for 2 weeks. Batch S2 was grown following the method described in Ref. [12]. Two high-pressure XRD runs on S1 and S2 were carried out at RT in beamline 16-BMD of HPCAT, the Advanced Photon Source (APS) in Debye-Scherrer geometry. Uranium $L_3$ edge RXES and PFY-XAS data at RT under various pressures were collected using samples from batch S2 at the HPCAT 16-IDD beamline. More details are in Supplementary Materials [13].

## Results and Discussion

### *High-pressure XRD*

Figure 1 shows waterfall plots of the $UTe_2$ (S1 and S2) XRD spectra under high pressure up to 30 GPa and 20 GPa, respectively. At near ambient pressure, $UTe_2$ spectra was identified with an orthorhombic structure and space group *Immm* [14] with no impurities. Obvious changes in the diffraction patterns, seen in Figure 1(b), between 5 and 7 GPa, indicate that a structural phase transition has occurred. The transition is complete at ~8 GPa, which agrees with results of Refs. [8, 9], with no further phase transitions observed up to 30

GPa. Small discrepancies in the transition pressure could be explained by our choice of different pressure transmitting media.

The reduced number of peaks in the XRD spectra of the high-pressure phase indicates an increased symmetry, which was identified as a body-centered tetragonal structure (space group *I4/mmm*) [15, 8, 9]. Following the analyses in Refs. [15, 8, and 9], the lattice parameters ($a$, $b$, $c$), and unit cell volume ($V$) of $UTe_2$ under pressure were calculated via Rietveld refinement of XRD peaks using the open-source software GSAS-II [16], shown in Figure 2. The atomic positions of U and Te atoms in the unit cell of the orthorhombic/tetragonal phase were also refined and to find the pressure dependence of the nearest uranium to uranium distance, $d_{U\text{-}U}$, in Figure 3. The initial unit cell volume ($V_o$) was refined from ambient pressure XRD spectra taken at RT from the same batch used in subsequent high pressure XRD measurements.

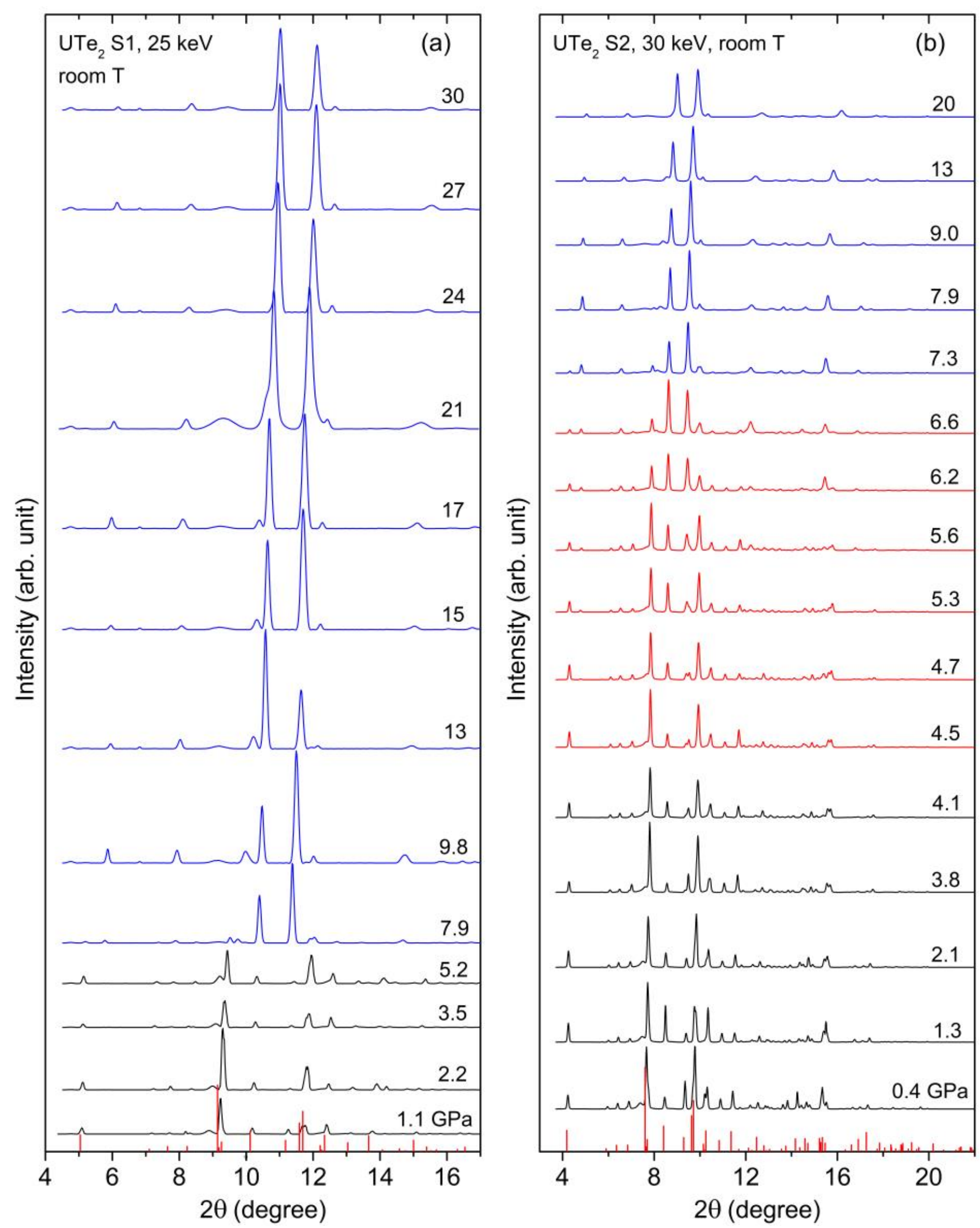

**Figure 1.** XRD spectra of $UTe_2$ measured at various pressures between 0.4 and 30 GPa. (a) XRD patterns of $UTe_2$ from batch S1; (b) XRD spectra of $UTe_2$ from batch S2, which show a structural phase transition in progress between 5 GPa to 7 GPa. The color scheme used to delineate the XRD spectra: Black – low-*P*

phase, blue – high-*P* phase, red – mixed low-*P* and high-*P* phases. Red vertical lines at the bottom of the figures indicate standard $UTe_2$ diffraction peaks based on Ref. [14].

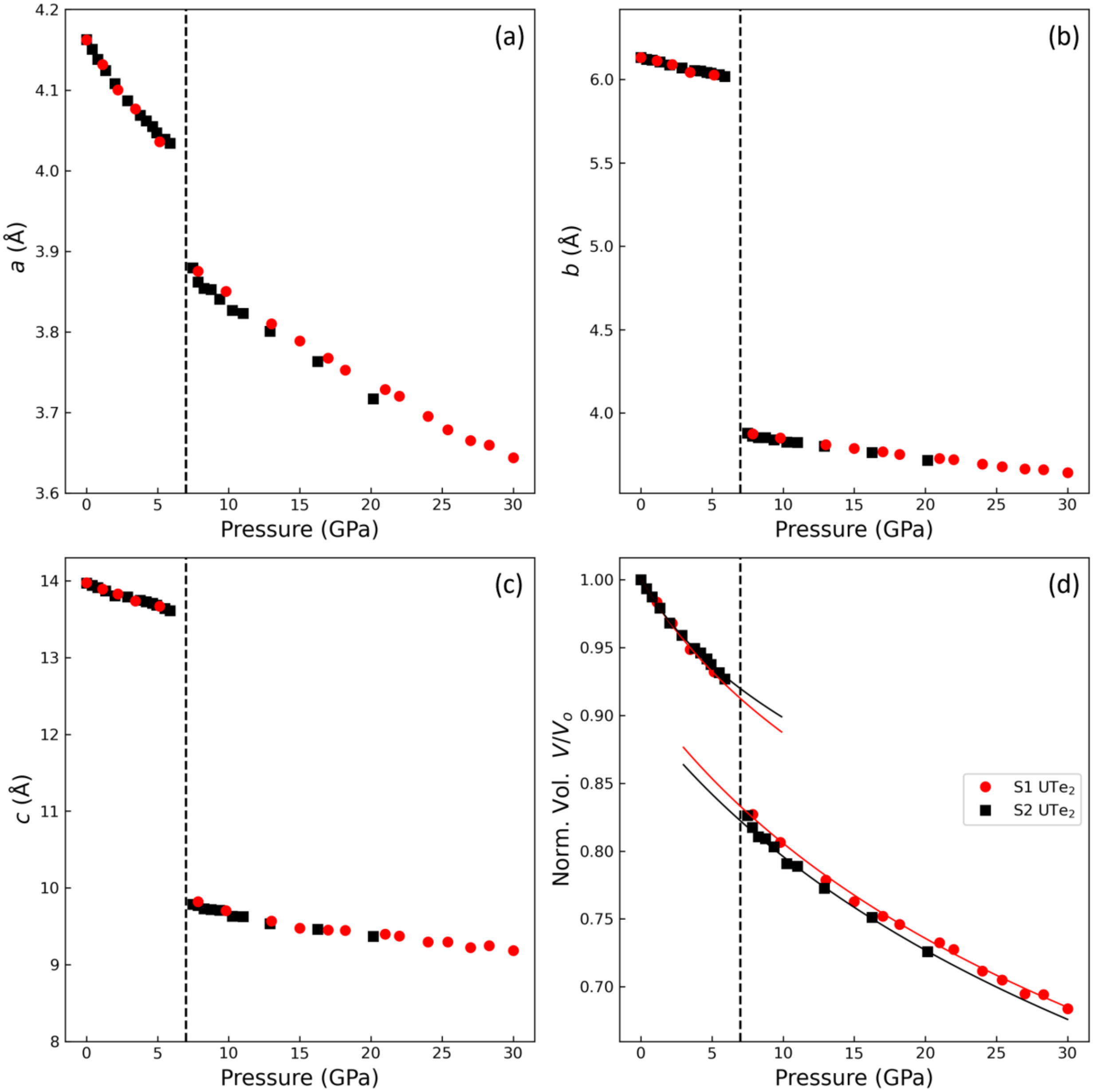

**Figure 2.** Lattice parameters for two samples of $UTe_2$ vs. pressure. (a - c): Lattice parameters *a*, *b*, and *c*. The current figure omits data points from the mixed phase for simplicity. (d): Normalized unit cell volume $(V/V_o)$ vs. increasing pressure with respective 3rd order Birch-Murnaghan equation of state fits for both phases. The tetragonal phase *V* was multiplied by 2 and normalized to the *Immm* $V_o$ to conserve the number of atoms across the structural phase transition.

Both samples of $UTe_2$ exhibit a relative volume decrease of $\Delta V/V_o$~11 % at the phase transition (Figure 2(d)). The *b* axis of the low-pressure orthorhombic lattice has the largest compression of ~33 % at the phase transition, followed by ~27% for the *c* axis and ~3% for the *a* axis. The bulk modulus ($K_0$) and differential bulk modulus ($dK_0/dP$) were calculated by fitting a 3rd order Birch-Murnaghan equation of state [8] to the *V* vs. *P* data,

$$P(V) = \frac{3}{2}K_0\left[\left(\frac{V_0}{V}\right)^{\frac{7}{3}} - \left(\frac{V_0}{V}\right)^{\frac{5}{3}}\right] \times \left\{1 + \frac{3}{4}\left(\frac{dK_0}{dP} - 4\right)\left[\left(\frac{V_0}{V}\right)^{\frac{2}{3}} - 1\right]\right\} \quad (1)$$

In Table 1, we list the fitting parameters of the 3rd order Birch-Murnaghan equation of state for our data shown in Figure 2(d).

Table 1. $V_0$, $K_0$, and $dK_0/dP$ of $UTe_2$ in the orthorhombic and tetragonal phases at RT.

| | Orthorhombic (*Immm*) | | | Tetragonal (*I4/mmm*) | | |
|---|---|---|---|---|---|---|
| | $V_0$ (Å$^3$) | $K_0$ (GPa) | $dK_0/dP$ | $V_0$ (Å$^3$) | $K_0$ (GPa) | $dK_0/dP$ |
| $UTe_2$ S1 | 356.73 | 60±5 | 6±3 | 163.31 | 62±13 | 3.6±0.7 |
| $UTe_2$ S2 | 356.63 | 57±4 | 6.2±2 | 160.31 | 67±2 | 3.3±0.4 |

The value of $K_0$ (avg. ~59 GPa) for the orthorhombic phase is comparable to that in Refs. [8] (~44 GPa) and [9] (~59 GPa). The high-pressure tetragonal $UTe_2$ phase has a stiffer bulk modulus of ~65 GPa, which is smaller than the value (~74 GPa) obtained in [8]. The bulk modulus of the orthorhombic phase is small, comparable to tellurium (65 GPa), suggesting $UTe_2$ is a very soft material whose physical properties are more responsive to changes in pressure. Interestingly, tellurium also undergoes a hexagonal to monoclinic structural transition at about 4 GPa, which is similar to the onset pressure for the structural phase transition for $UTe_2$ [17].

The change in $d_{U\text{-}U}$ with pressure was also calculated (shown in Figure 3) for the two structures. Notably, $d_{U\text{-}U}$ increased at the structural phase transition, from ~3.7 Å to ~3.9 Å, comparable to the values reported in Refs. [9] and [8]. Even at 30 GPa, the $d_{U\text{-}U}$ is still above the Hill limit of uranium compounds (3.4 - 3.6 Å).

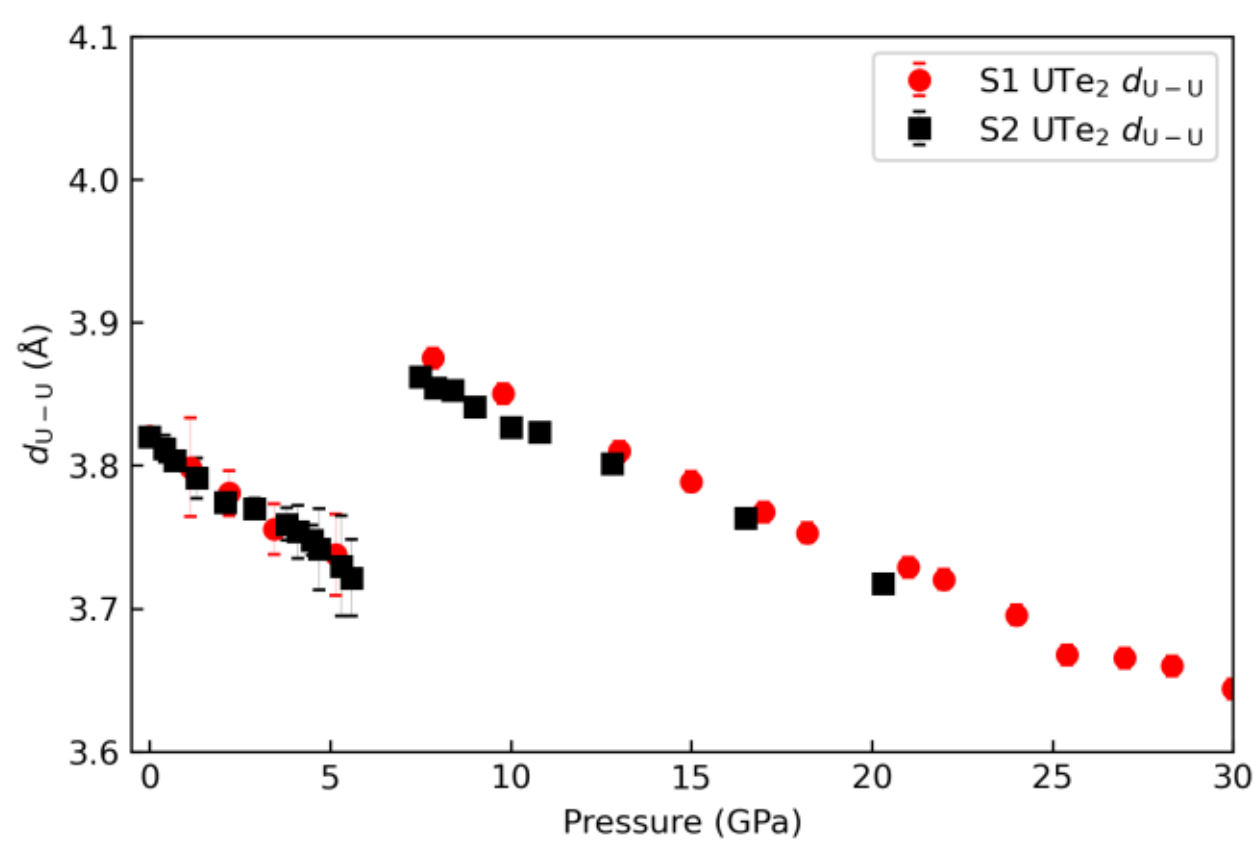

**Figure 3.** The change in $d_{U\text{-}U}$ in $UTe_2$ with pressure. For both samples, $d_{U\text{-}U}$ increases at the structural phase transition from *Immm* to *I4/mmm* and steadily decreases with increasing pressure, where $d_{U\text{-}U}$ in the high-pressure tetragonal phase at ~20 GPa equals the value in the lower pressure orthorhombic phase at 6 GPa. The vertical bars represent the $d_{U\text{-}U}$ measurement uncertainty. See Ref. [13] for details.

Using the CALYPSO software package and Vienna ab initio simulation package (VASP), Hu *et al.* predicted that the *Immm* to *I4/mmm* structural transition would occur at 4 GPa accompanied by a ~12% volume collapse at the transition, when implementing the Perdew-Burke-Ernzerhof (PBE) + Hubbard $U$ method with $U$ = 1 or 2 eV [15]. This value is quite comparable to our experimental results of ~11 % decrease in volume. The denser high pressure *I4/mmm* phase is achieved through a transition from a U-Te eight-coordinated distorted square antiprism [8] to a ten-coordinated "bicapped cube", in which lone pair electrons of Te participate more in forming covalent bonding with U atoms, according to the calculation of the projected two-dimensional electron localization functions [15].

Crystallographic and electronic structures are intimately interrelated, which underlines the importance of studying the pressure induced structural phase transition in $UTe_2$, since it can provide insights into how 5*f*-electrons are involved in the bonding. When more 5*f*-electrons participate in bonding in actinides it is typical to observe lower crystallographic symmetry [18, 19]. With respect to the behavior of $UTe_2$ under pressure, the $d_{U-U}$ in $UTe_2$ (~3.8 Å) at ambient pressure is larger than the Hill limit (~3.5 Å), a "rule-of-thumb" parameter representing the boundary between localization and delocalization of 5*f*-electrons for U compounds. Thus, we should expect magnetic order in $UTe_2$ instead of the superconductivity [20] observed in $UTe_2$ at ambient pressure. Strangely, the increase in $d_{U-U}$ at ~5 GPa indicates the 5*f*-electrons of $UTe_2$ become more localized under pressure, which is consistent with the transition from the lower symmetry orthorhombic phase to the higher symmetry tetragonal phase, if the correspondence between 5*f*-localization and higher crystal structure symmetry is applicable to $UTe_2$. The scenario of increased localization of 5*f*-electrons in $UTe_2$ under pressure is supported by measurements of the electrical resistivity as a function of temperature reported in Ref. [9]. The measurements reveal the occurrence of superconductivity above 6 GPa with an upper critical field lower than the Pauli limit and Fermi liquid behavior in the normal state electrical resistivity $\rho(T)$ with a small coefficient of the $T^2$ term, indicating that the electronic state of tetragonal $UTe_2$ is weakly correlated [9].

Huston *et al.* [8] pointed out that in other uranium chalcogenides such as USe and UTe which have a ferromagnetic ground state and a $d_{U-U}$ that exceeds the Hill limit, pressure also drives a transition from lower to higher structural symmetry [21, 8]. It is interesting to note, regardless of whether the ground state is superconducting as in $UTe_2$, or ferromagnetic as in USe and UTe, uranium chalcogenides seem inclined to transform into a higher symmetry phase at high pressure. In the first-principles study of $UTe_2$ by Hu *et al.*, it was shown that covalent bonding between lone pair electrons of Te and U atoms forms in the high-pressure tetragonal phase [15]. Perhaps this covalent bonding localizes 5*f*-electrons, thus inducing the higher symmetry tetragonal phase and increasing the bulk modulus appreciably since a three-dimensional network of covalent bonds (e.g., diamond and quartz) is usually quite rigid (although lattice stiffening is common in many kinds of pressure induced phase changes) [22].

*High-pressure PFY-XAS*

The electronic configuration of U in $UTe_2$ at ambient or high pressures has been studied using a variety of x-ray spectroscopy techniques [10, 11, 23]. Here we explore the PFY-XAS of the U $L_3$ absorption from $2p_{3/2}$ to an unoccupied 6*d* orbital and the RXES of the U $L_{\alpha1}$ emission from $3d_{5/2}$ to $2p_{3/2}$ to study the electronic state of $UTe_2$. PFY-XAS measurements were taken by exciting the U $L_3$ absorption edge with incident energy between 17.140 - 17.210 keV, while detecting emission at the U $L_{\alpha1}$ = 13.614 keV line. Measurements were performed at RT at various pressures in DACs. As can be seen in Figure 4, our PFY-XAS spectra for $UTe_2$ show the white-line (main peak) from the $3d_{5/2}$ to $2p_{3/2}$ transition, a step-function like background associated with the excitation into the continuum above the 6*d* states, and a small pre-edge shoulder at 6-7 eV below the white-line. The shoulder can be attributed to a pre-edge quadrupole transition from $2p_{3/2}$ to 5*f* [24], which is usually smaller compared to the dipole allowed white-line. A similar feature on the left side of the white-line was also found in $U^{4+}$, $U^{5+}$, and $U^{6+}$ compounds, with a separation of 5-7 eV [24, 25]. PFY-XAS spectra was normalized to the post-edge step above 17.170 keV due to the excitation into the continuum using the open-sourced Larch XAS analysis software [26], while additional post white-line flattening was applied to the data at $P$ = 1.8 GPa due to synchrotron beam current instability that occurred during measurement (subsequent measurements did not have this issue). Spectra were fitted with a combination of an arctangent step function for data above 17.170 keV and a Gaussian function for the white-line, using the method outlined by Ref. [10]. The model was adjusted by varying all parameters with initial guesses for the step function width set to the uranium $L_3$ core-hole lifetime (~3.9 eV) and the step function position set by the first derivative of the interpolated PFY-XAS data curve. A detailed description and a figure showing all deconvoluted PFY-XAS spectra are presented in Supplementary Materials [13].

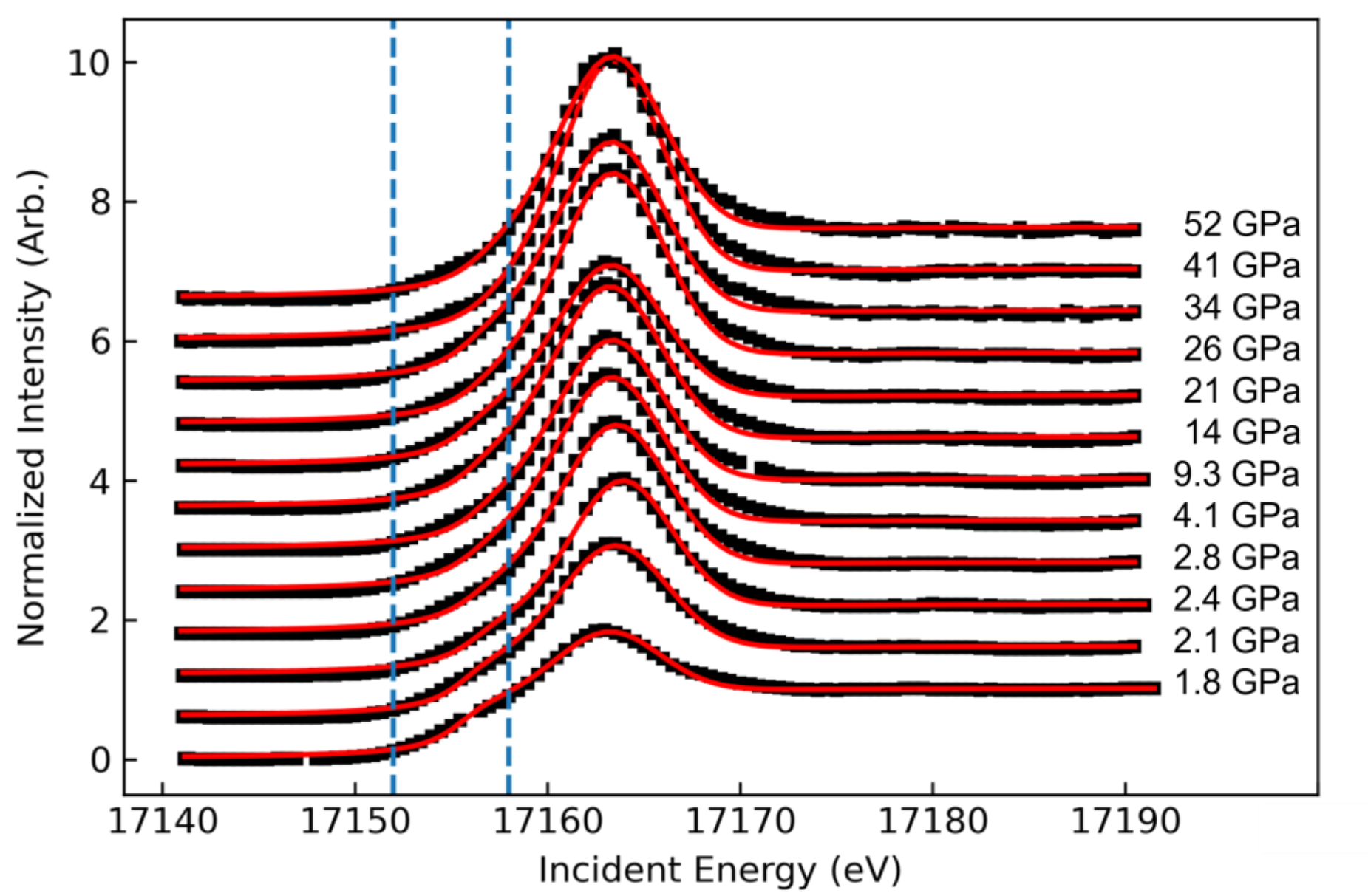

**Figure 4.** PFY-XAS spectra of $UTe_2$ at different pressures. Curves are offset vertically for clarity. The emission intensity is normalized to the incident intensity. The two blue dashed lines enclose the region where the pre-edge quadrupole resonance occurs. The red curves are fits to the data which are represented by solid black squares (see Supplementary Materials [13] for more discussion).

Figure 5 shows the change in the white-line peak position under pressure, which was found by plotting the $E_i$ of the peak maximum (see [13]). The white-line position blueshifts by ~0.7 eV up to 2.4 GPa. However, subsequent increases in pressure results in a redshift white-line up to 4.1 GPa, after which there is little change in peak position up to 52 GPa. In previous studies of uranium intermetallics, changes in the U $L_3$ edge white-line position were attributed to a change in U valence (or 5*f*-electron count) [10] or a change in the degree of delocalization of the 5*f*-electrons [7]. Although there are contradictory conclusions on the 5*f*-electron count of $UTe_2$ (closer to 3 [27, 11] vs closer to 2 [10]) at ambient pressure, $UTe_2$ is believed to have an intermediate valence. Thomas *et al*. reported a small increase in U valence towards 4+ above 1.25 GPa [10] in $UTe_2$ under pressure. Using U $M_{4,5}$ edge XANES, Wilhelm *et al*. [11] observed an initial increase in valence towards 4+ with pressure up to ~2 GPa, followed by a decrease in valence back to 3+ with pressure. The white-line blueshift below 2.4 GPa can be ascribed to a pressure-induced delocalization of 5*f*-electrons, or a transfer of spectral weight from $5f^3$ to $5f^2$ configuration, or a combination of both. Considering that the $d_{U\text{-}U}$ is larger than the Hill limit, delocalization can occur through the hybridization between the conduction band and 5*f* orbitals. Between 2.4 and 4.1 GPa, the observed redshift of the $L_3$ white-line indicates a reverse trend: 5*f*-electrons become more localized or there is an increase in the occupancy of the $5f^3$ configuration. The stable white-line position between 4.1 and 52 GPa reveals a 5*f* configuration with a stable degree of localization or a stable 5*f* configuration. The multiple origins of the white-line shift and the limited resolution of PFY-XAS data prevented an estimation of the 5*f* occupancy using standard first x-ray sum rules [28], so an estimation of the 5*f* occupancy was determined from deconvoluting RXES data instead, as described below.

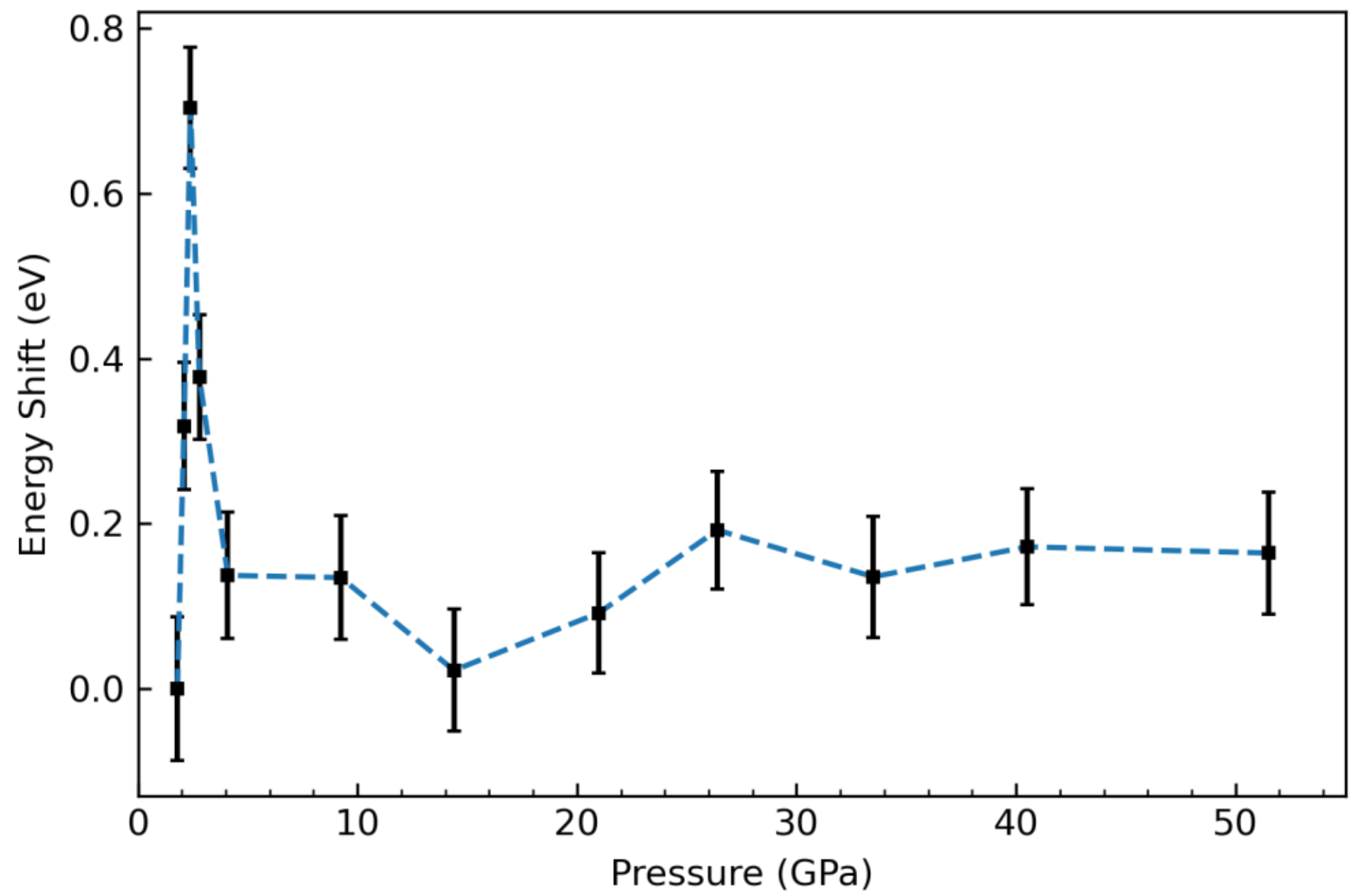

**Figure 5.** Change in $UTe_2$ S1 U $L_3$ edge white-line position for $UTe_2$ with pressure relative to its position at 1.8 GPa. The vertical bars indicate the fitting error determined by estimated standard deviation [13].

*High pressure RXES*

Representative RXES spectra of normalized $L_{\alpha 1}$ emission of $UTe_2$ at 2.1 GPa are shown in Figure 6 as a function of incident x-ray energy ($E_i$) and transferred energy $E_t \equiv E_i - E_e$ where $E_e$ is the emission energy. Even before deconvoluting configuration peaks there is a distinguishable doublet in the spectra at $E_i$ below the white-line. Peak positions stay nearly constant with $E_t$ when $E_i$ is lower than a threshold $E_T = 17.170$ keV above which the non-resonant x-ray emission becomes dominant. This is characteristic of the resonant emission from $3d_{5/2}$ to $2p_{3/2}$ [5]. When $E_i$ is higher than $E_T$, the excitation into the continuum overwhelms the excitation into the 6*d* states, and the emission spectra as a function of $E_t$ shift linearly with $E_i$ [5, 6], as indicated by curves with $E_i > E_T$ in Figure 6. The multiple peaks in the RXES emission spectra indicate multiconfigurational 5*f* states and agrees with the PFY-XAS results. This suggests $UTe_2$ falls into a similar grouping of intermediate valence materials like $SmB_6$, $YbAl_3$ [29], and $UPd_2Al_3$ [30].

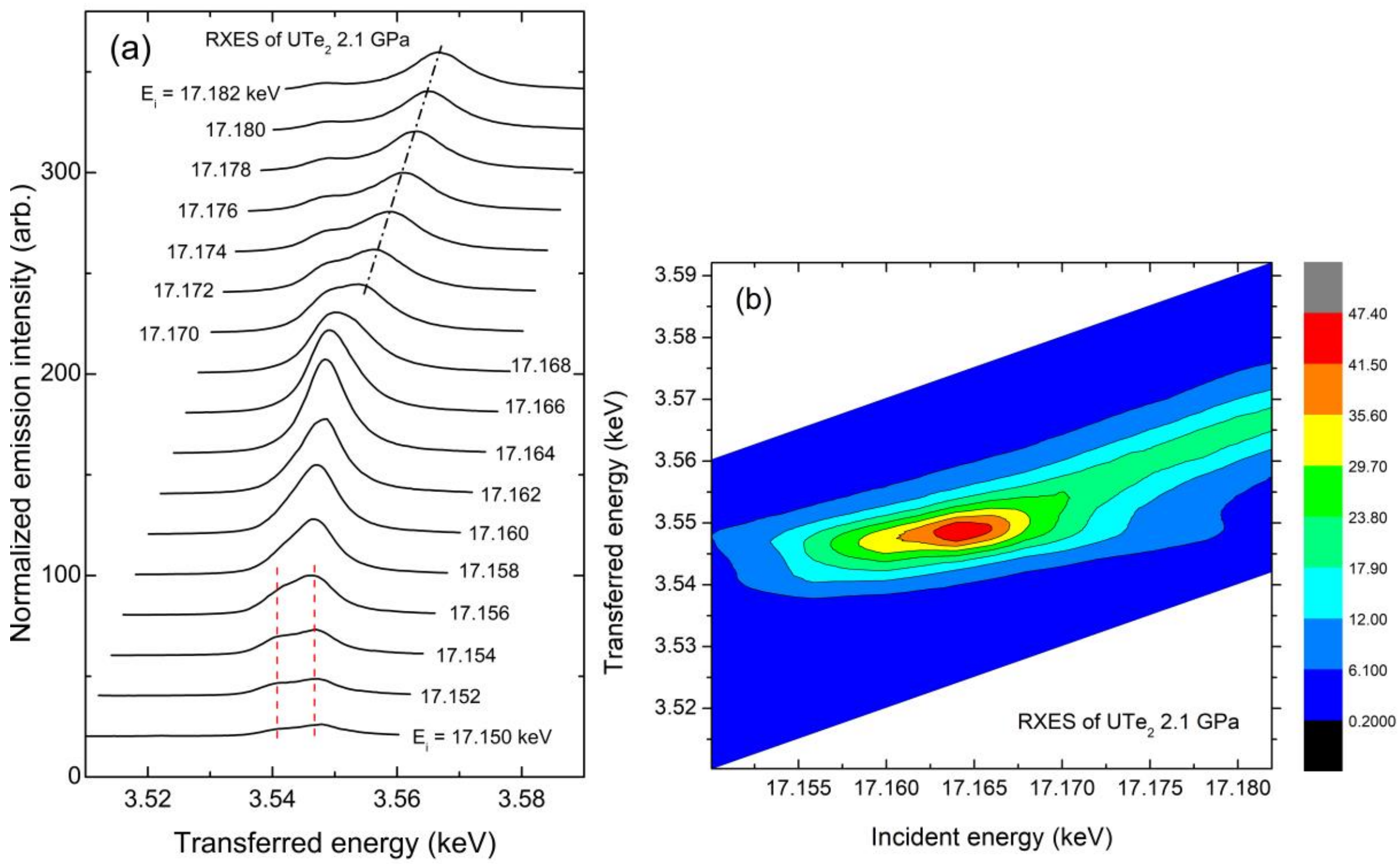

**Figure 6.** RXES of $UTe_2$ S1 at 2.1 GPa and RT. The red dashed lines indicate the two 5*f* configurations and the black dash-dotted line indicate the shift in the non-resonant emission. The colors in the vertical scale bar of (b) show the normalized emission intensity. When $E_i \geq 17.170$ keV, a clear linear correlation can be seen between $E_t$ at the emission maximum and $E_i$, indicating emission is from non-resonant excitation to the continuum.

The 5*f* configuration weights ($5f^3$, $5f^2$) were calculated using the methodology in Ref. [6] for $UCd_{11}$ and $PuSb_2$. Since the total number of unoccupied 6*d* states is almost fixed, the relative excitation amplitude should be proportional to the occupancy in each 5*f* configuration [6]. The 5*f*-electron occupancy can then be calculated from the weighted sum of the integrated multiconfiguration peaks that are deconvoluted from the RXES spectra. Fortunately, unlike some other U intermetallics (e.g., $UCd_{11}$ [6], $URu_2Si_2$ [25]), our RXES data at energy below $E_T$ clearly exhibit separate $f^3$ and $f^2$ peaks from the emission signal which helped identify the position of each configuration. Deconvolution of RXES data was accomplished via a non-linear least square fitting python module (LMFIT) [31], by using two skewed Lorentzian functions corresponding to the $f^3$ and $f^2$ peaks, and a third skewed Lorentzian function corresponding to the fluorescence peak (FP) when $E_i$ approached $E_T$ of 17.170 keV to account for the electrons excited to the continuum.

An example multiconfiguration peak deconvolution at 2.4 GPa is shown in Figure 7(a) where green and red peaks correspond to the $5f^3$ and $5f^2$ configurations, respectively, and the purple dashed FP corresponds to the non-resonant emission at selected $E_i$ at 2.4 GPa. The areas under the curves are integrated, to calculate the respective peak coefficients, shown in Figure 7(b). The resulting normalized peak coefficient data were then fitted to an associated Lorentzian (for $5f^3$ and $5f^2$ peaks) and integrated to calculate the configuration weight at each pressure. A more detailed summary for peak deconvolution and error analysis is discussed in Supplementary Materials [13].

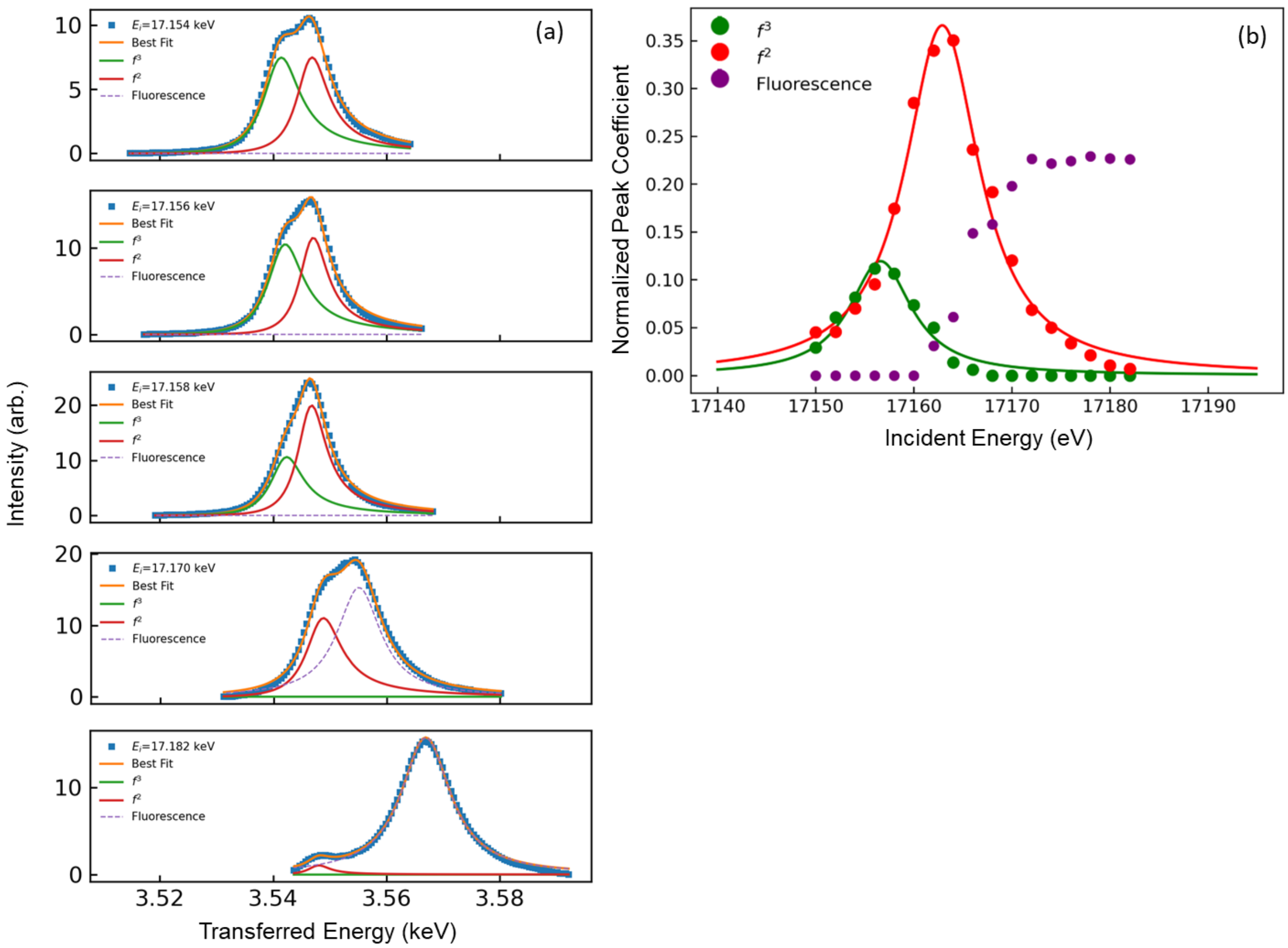

**Figure 7.** (a) Multiconfiguration peak deconvolution for $L_{\alpha 1}$ RXES at 2.4 GPa where green and red peaks correspond to the $5f^3$ and $5f^2$ peaks, respectively, and the purple dashed FP corresponds to the non-resonant emission. The FP dominates over $5f^3$ and $5f^2$ peaks when $E_i \geq 17170$ eV. (b) Normalized peak coefficient data plotted vs $E_i$. The fitted Lorentzians were integrated to calculate the configuration weights.

The $5f$ occupation ($n_f$) and $f^3$ and $f^2$ configuration fractions results are shown in Figure 8. At low-pressure, an initial $5f$ occupancy $n_f$ = ~2.26 (valence of 3.74) at 1.8 GPa decreases with pressure until a minimum at ~2.8 GPa where $n_f$ ~2.21 (valence of 3.79). Afterwards, $n_f$ increases with pressure until ~15 GPa where a plateau appears at $n_f$ ~2.24 (valence of 3.76) and persists up to 52 GPa, indicating that the $UTe_2$ valence stays mostly constant at high pressures. The nonmonotonic valence change preceding the pressure-induced phase transition (5 GPa) qualitatively agrees with Ref. [11] which claims a valence maximum at about 2 GPa. The movement of the valence towards 3+ seems to be associated with the phase transition to the tetragonal structure which has more room for U atoms due to the larger $d_{U\text{-}U}$ (see Figure 3). The increased volume can more easily accommodate the $U^{3+}$ ion, which has a larger ionic radius than the $U^{4+}$ ion. Stability of the valence above 15 GPa appears to be consistent with PFY-XAS data which exhibit little energy shift in the white-line position with pressure (see Figure 5). Although the PFY-XAS white-line shift can be due to both the change in $n_f$ and in the localization of $5f$-electrons [32], we do see some similarities between the PFY-XAS and RXES data, as manifested in the mirrored curves in Figures 5 and 8. The discrepancy between the two curves reflects the extra contribution to the PFY-XAS spectra from the change in the degree of itinerancy of $5f$-electrons under pressure.

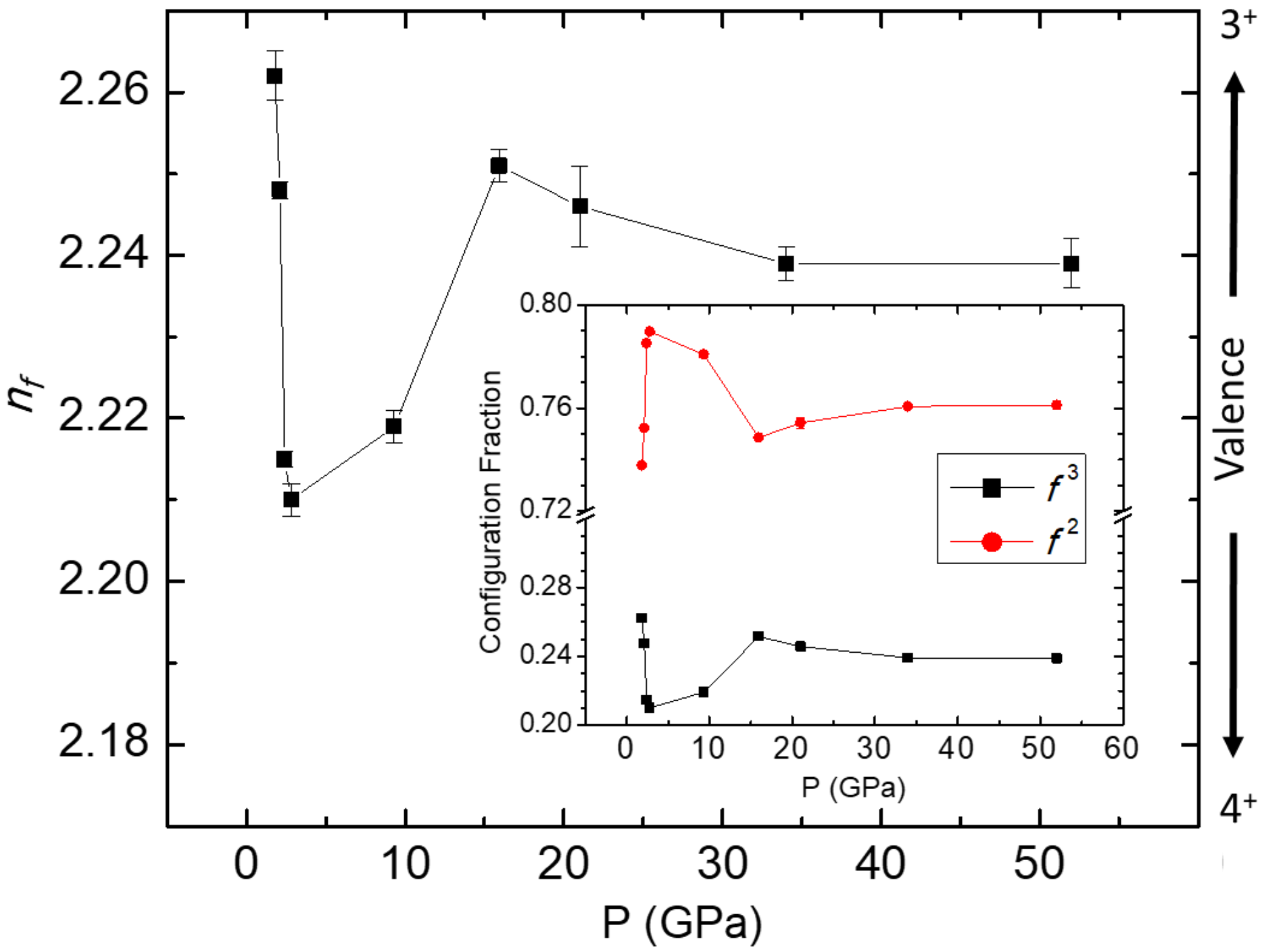

**Figure 8.** The calculated 5$f$ occupation $n_f$, and configuration fractions of $f^3$ and $f^2$ (inset) as a function of pressure for $UTe_2$ at RT. The vertical bars indicate the estimated standard deviation (refer to Supplementary Materials [13]).

In a Kondo lattice, magnetic ordering of the lanthanide or actinide magnetic moments is mediated by conduction electrons whose spins interact with the lanthanide or actinide magnetic moments via the Ruderman-Kittel-Kasuya-Yosida RKKY interaction that varies to second order in the intra-atomic exchange interaction parameter $J_{ex} = -V_H^2/|\varepsilon_f|$, where $V_H$ represents the strength of hybridization between the localized $f$- and conduction electron-states and $\varepsilon_f$ is the energy separating the localized $f$-electron state and the Fermi level $E_F$. In $UTe_2$, with decreasing pressure in the lower pressure region (1.8 – 2.8 GPa), the valence changes towards 3+, coincident with the rapid suppression of the Néel temperature of the AFM phase [10, 33], followed by the appearance of two unconventional superconducting phases near ambient pressure and a corresponding heavy fermion normal state, reflected in the $T$-dependence of the electrical resistivity. This reveals the location of an AFM QCP in the $UTe_2$ Doniach phase diagram associated with the complete Kondo screening of the U magnetic moments induced by decreasing pressure. It would be interesting to determine if this is accompanied by a simultaneous valence change in the same pressure region. Additionally, it implies $|J_{ex}|$ increases with decreasing pressure, which is unusual because we expect $|J_{ex}|$ to decrease with decreasing pressure due to a decrease in $V_H$ with decreasing pressure. Similarly, it infers $|\varepsilon_f|$ decreases with decreasing pressure at a higher rate than that of $V_H$. This is similar to the scenario suggested for the relationship between the hidden order phase and the antiferromagnetic phase in the heavy fermion compound $URu_2Si_2$ [34]. In that case, pressure or chemical pressure via substitution of the smaller Fe atom for Ru drives the transition in $URu_2Si_2$ from the nonmagnetic hidden order phase, which exhibits coexisting unconventional superconductivity, to the AFM phase in the Doniach phase diagram due to a decrease in $|J_{ex}|$ with pressure [34].

## Concluding Remarks

We studied the electronic and structural properties of $UTe_2$ via RXES up to 52 GPa and XRD up to 30 GPa, to higher pressures than previously [10, 11], and mapped the valence across the pressure induced structural phase transition between 5 and 7 GPa from *Immm* (orthorhombic) to *I4/mmm* (body centered tetragonal). Due to the higher energy resolution of the RXES technique compared to traditional XANES, we were able to calculate the change of the U valence with pressure. Observations indicate that $UTe_2$ as an *f*-electron system changes from a lower to a higher symmetry crystal structure under high pressure. This indicates that in the high-pressure tetragonal phase, 5*f*-electrons become more localized and less itinerant as one would expect from the viewpoint of a pressure induced increase in $d_{\mathrm{U\text{-}U}}$. The $d_{\mathrm{U\text{-}U}}$ over the whole investigated pressure range (1.8 to 30 GPa) is larger than the Hill limit for U compounds, so magnetic ordering due to RKKY interactions could be anticipated. In reality, in this regime correlated electron phenomena such as unconventional superconductivity and heavy fermion behavior emerge. This implies that hybridization ($V_{\mathrm{H}}$) between localized 5*f*- and conduction electron-states and on-site Coulomb repulsion (*U*) between 5*f*-electrons should play an important role in determining the electronic and magnetic properties of $UTe_2$ [8], like it does in other heavy fermion systems, contributing to the failure of using the Hill limit to categorize the degree of 5*f* itinerancy in $UTe_2$.

At RT, RXES data show $UTe_2$ has a valence of ~3.74 at 1.8 GPa which changes to 3.79 at 2.8 GPa and then back to a value slightly smaller than that at 1.8 GPa. Interestingly, the change in valence is accompanied by the structural phase transition at 5 - 7 GPa, suggesting a correlation between the U valence and crystal structure of $UTe_2$. At pressures higher than 15 GPa, the U valence does not change with pressure. PFY-XAS data also revealed a nonmonotonic change in the white-line position with pressure; however, due to complications associated with different mechanisms that lead to the shift in white-line position, it is difficult to reach a more definitive conclusion. It is worthwhile to point out that our PFY-XAS and RXES measurements were done at RT whereas other results regarding the $UTe_2$ valence were based on spectroscopic measurements at low temperature in Ref. [27] (20 K), Ref. [10] (1.7 K), and Ref. [11] (2.7 K). It is known that for 4*f* systems with a small Kondo temperature, $n_f$ depends strongly on temperature (e.g., $YbAgCu_4$, see [35]). Our RXES results demonstrate well-separated resonant emissions from $5f^{3}$ and $5f^{2}$ configurations, which illustrate the advantage of using RXES in determining $UTe_2$ valence. Extending RXES measurements to lower temperature, especially near the superconducting or magnetic ordering temperatures at high pressures, would reveal whether the U valence changes when $UTe_2$ undergoes these electronic and magnetic transitions.

## Acknowledgments

Research at University of California, San Diego was supported by the National Nuclear Security Administration (NNSA) under the Stewardship Science Academic Alliance Program through the US DOE under Grant DE-NA0004086, and the US Department of Energy (DOE) Basic Energy Sciences under Grant DE-FG02-04ER46105. This work was sponsored in part by the UC San Diego Materials Research Science and Engineering Center (UCSD MRSEC), supported by the National Science Foundation (NSF) under Grant DMR-2011924. Research at the University of Illinois Chicago was supported by the NNSA (DE-NA-0003975, CDAC) and NSF (DMR-2119308). Research at Florida State University and the National High Magnetic Field Laboratory was supported by NSF Cooperative Agreement DMR-2128556, and the DOE. Portions of this work were performed at HPCAT (Sector 16), Advanced Photon Source (APS), Argonne National Laboratory. HPCAT operations are supported by DOE-NNSA's Office of Experimental Sciences. The Advanced Photon Source is a U.S. Department of Energy (DOE) Office of Science User Facility operated for the DOE Office of Science by Argonne National Laboratory under Contract No. DE-AC02-06CH11357. It is a pleasure to acknowledge informative discussions with Professor Andrea Severing, Professor Liu Hao Tjeng, Andrea Marino, and Denise Christovam. We thank Curtis Kenney-Benson for assistance with x-ray measurements at HPCAT.